\documentclass[fleqn,usenatbib]{mnras}
\usepackage[utf8]{inputenc}
\usepackage[varg]{txfonts}
\usepackage{color} 
\usepackage[english]{babel}
\usepackage{fontenc,amssymb}
\usepackage{graphicx}
\usepackage{subfigure}

\newcommand{\Eq}[1]{Eq.~\ref{#1}}

\newcommand{\Fig}[1]{Fig.~\ref{#1}}

\newcommand{\Nciv}{19}
\newcommand{\mgiimean}{0.99}

\usepackage[normalem]{ulem}
\newcommand{\CIV}{\hbox{{\rm C}{\sc \,iv}}}

\newcommand{\OII}{\hbox{{\rm O}{\sc \,ii}}}

\newcommand{\MgII}{\hbox{{\rm Mg}{\sc \,ii}}}

\newcommand{\flux}{erg\,s$^{-1}$\,cm$^{-2}$}

\newcommand{\mpy}{\hbox{M$_{\odot}$\,yr$^{-1}$}}

\newcommand{\Mh}{M_{\rm h}}
\newcommand{\Rvir}{R_{\rm vir}}
\newcommand{\fc}{f_{\rm c}}
\newcommand{\kms}{\hbox{km~s$^{-1}$}}
\newcommand{\Mpcm}{\hbox{Mpc$^{-1}$}}
\newcommand{\kpc}{\hbox{kpc}}

\newcommand{\fCIV}{1.36$\pm0.57$}
\newcommand{\fMgII}{1.67$\pm0.15$}

\title[A study of \MgII\ and \CIV\ in MEGAFLOW]{MusE GAs Flow and Wind (MEGAFLOW) VI. 
A study of \CIV\ and \MgII\ absorbing gas surrounding [\OII] emitting galaxies~\thanks{Based on observations made with ESO Telescopes at the La Silla Paranal Observatory under programme IDs 
094.A-0211,
095.A-0365,
096.A-0609, %UVES
096.A-0164,
097.A-0138,
097.A-0144, %UVES
098.A-0216,
098.A-0310, %UVES
099.A-0059,
293.A-5038 %UVES,
}}
\author[I. Schroetter et al.]{
Ilane~Schroetter,$^{1,5}$\thanks{E-mail: ilane.schroetter@gmail.com}
Nicolas~F.~Bouch\'e,$^{2}$ 
Johannes~Zabl,$^{2,9}$
Hadi~Rahmani,$^{1,7}$
Martin~Wendt,$^{3,4}$
\newauthor
Sowgat~Muzahid,$^{8,4}$
Thierry~Contini,$^{5}$
Joop~Schaye,$^{6}$
Kasper~B.~Schmidt,$^{4}$
Lutz~Wisotzki~$^{4}$
\\
$^{1}$ GEPI, Observatoire de Paris, PSL Université, CNRS,  5 Place Jules Janssen, 92190 Meudon, France\\
$^{2}$ Univ Lyon, Univ Lyon1, Ens de Lyon, CNRS, Centre de Recherche Astrophysique de Lyon UMR5574, F-69230 Saint-Genis-Laval, France\\
$^{3}$ Institut f\"ur Physik und Astronomie, Universit\"at Potsdam, Karl-Liebknecht-Str. 24/25, 14476 Golm, Germany \\
$^{4}$ Leibniz-Institut für Astrophysik Potsdam, An der Sternwarte 16, D-14482 Potsdam, Germany \\
$^{5}$ Institut de Recherche en Astrophysique et Plan\'etologie (IRAP), Universit\'e de Toulouse, CNRS, UPS, F-31400 Toulouse, France\\
$^{6}$ Leiden Observatory, Leiden University, PO Box 9513, 2300 RA Leiden, The Netherlands \\
$^{7}$ School of Astronomy, Institute for Research in Fundamental Sciences (IPM), PO Box 19395-5531 Tehran, Iran\\
$^{8}$ IUCAA, Post Bag 04, Ganeshkhind, Pune, India-411007  \\
$^{9}$ Institute for Computational Astrophysics and Department of Astronomy \& Physics, Saint Mary’s University, 923 Robie Street, Halifax, Nova Scotia, B3H 3C3, Canada\\
}

\begin{document}
\label{firstpage}
\pagerange{\pageref{firstpage}--\pageref{lastpage}}
\maketitle
\begin{abstract}
Using the MEGAFLOW survey, which consists of a combination of MUSE and UVES observations of 22 quasar fields selected to contain strong \MgII\ absorbers, 
we measure covering fractions of \CIV\ and \MgII\ as a function of  impact parameter $b$
using a novel Bayesian logistic regression method on unbinned data, appropriate for small samples. 
{We also analyse how the \CIV\ and \MgII\ covering fractions evolve with redshift.}
In the MUSE data, we found {215}  $z=1-1.5$ [\OII] emitters with fluxes $>10^{-17}$ \flux\ and within 250 kpc of quasar sight-lines.
Over this redshift path $z=1-1.5$, we have  19 ({32})  \CIV\ (\MgII)  absorption systems with rest-frame equivalent width (REW) $W_r>$0.05\AA\ associated with at least one [\OII] emitter.
The  covering fractions of $z\approx1.2$ \CIV\  (\MgII) absorbers with mean $W_r\approx$0.7\AA\ (1.0\AA),   exceeds 50\% within 23$^{+62}_{-16}$ (46$^{+{18}}_{-13}$) kpc.
{Together with published studies, our results suggest that the covering fraction  of \CIV\  (\MgII) 
becomes larger (smaller) with time, respectively}.
{For absorption systems that have \CIV{} but not \MgII{}, we find in {{73}\%} of the cases no [\OII] counterpart.
This may indicate that the \CIV{} }
come from the intergalactic medium (IGM), i.e. beyond 250 kpc, or that it is associated with lower-mass or quiescent galaxies.

\end{abstract}
\section{Introduction}
\label{sec:intro}

The surroundings of galaxies, the intergalactic medium (IGM) or the very close environment around galaxies: the circum-galactic medium (CGM), are known to be filled with gas in various phases. 
This {CGM is composed by both} ionized {and} neutral {gas}. 
{For instance, \MgII{} is probing the {cool} ionized T$\sim10^4$~K temperature gas whereas \CIV{} probes {warm} T$\sim10^5$K gas \citep[see the CGM review of][]{TumlinsonJ_17a}.}
Because the gas usually has low density compared to the gas in galaxies, it is very difficult to observe directly. 
Understanding the origin of the gas surrounding galaxies is one of the keys to drawing a precise picture of how galaxies form and evolve. 

{In order to study the structure of the gas surrounding galaxies, the covering fraction ($f_{\rm c}$) as a function of impact parameter is a key diagnostic tool. Indeed, in numerical simulations, column density profiles
\citep[e.g.][]{HummelsC_13a} and covering fractions $f_c$
\citep[e.g.][]{BordoloiR_14b,LiangC_16a,LiF_21a} of low and high ionization elements are sensitive to feedback implementations and to added physical processes such as cosmic rays (CRs).
Low ionization \MgII\ ($\lambda \lambda 2796, 2803$) is a useful probe of the cool ($T\sim10^4$~K) CGM
%at $z<1.5$
because it is a doublet accessible over a wide {range} of redshifts with optical spectrographs.
Since the pioneering work of 
\citet{BergeronJ_86a} and \citet{SteidelC_92a}, numerous groups have constrained the covering fraction $f_{\rm c,\MgII}$ with ever increasing samples \citep[e.g.][]{ChenHW_10a,NielsenN_13a,DuttaR_20a, HuangY_21a} and perhaps the largest sample is \citet{LanT_19b} who used 15,000 galaxy-quasar pairs.
}

{
However, concerning \CIV\ ($\lambda \lambda 1548, 1550$),  studies of $f_{\rm c} (\CIV)$ are often limited to relatively small samples because of UV coverage required or {the} need to find the host galaxies at $z>2$ \citep[e.g][]{FoxA_07,FoxA_16}.
Among the earlier \CIV\ studies, there is the $z<1$  \citet{ChenH_01} study. Since the availability of the Cosmic Origin Spectrograph (COS) onboard the {\it Hubble} Space Telescope, we note the $z<0.1$ work of \citet{BordoloiR_14b} and \citet{BurchettJ_16a} on COS quasar sight-lines.  %\citet{Bish2021} 
In addition, there are other constraints on the \CIV\ CGM with quasar-quasar pairs  \citep{ProchaskaJ_14a,LandoniM_16}, with lensed
quasars \citep{RauchM_01} and with statistical clustering \citep[e.g.][]{SchayeJ_03a,TurnerM_14} which can be compared to hydrodynamical simulations \citep[e.g.][]{TurnerM_17}.
While these agree that the \CIV\ gas is mainly located outside their host galaxies,  there is a lack of consensus on the origin and properties of this highly ionized gas.  
 As a result the origin of this highly ionized gas, \CIV, {and its distribution around galaxies}, is thus still debated. 
}

In this paper, we aim 
{to address the question of how \CIV\ is located around galaxies compared to \MgII\ by searching for \CIV\ absorption systems and  their possible galaxy counterparts   
} by studying the \CIV\ and \MgII\ covering fractions using the
MEGAFLOW survey \citep{SchroetterI_16a, SchroetterI_19a,ZablJ_19a,ZablJ_20a}.
{The \CIV\ and \MgII\ covering fractions are obtained from a novel Bayesian logistic fit to the unbinned data}. {This method is designed to go beyond the limitations of using binned statistics and the Bayesian fit provides a built-in robustness to outliers.} 

The paper is organized as follows. We present  in ~\S~\ref{sec:data} the data and the independent detection of absorption systems and star-forming galaxies. 
We then combine those detections in order to identify possible galaxy counterparts to the absorbers. 
In \S~\ref{sec:covering} we focus on constraining covering fractions of \CIV\ and \MgII, we present our results in \S~\ref{sec:results} and our conclusions in \S~\ref{sec:conclusion}.
Throughout, we use a 737 cosmology ($H_0=70$~\kms~\Mpcm, $\Omega_{\rm m}=0.3$, and $\Omega_{\Lambda} = 0.7$) and a \citet{ChabrierG_03a} stellar Initial Mass Function (IMF).
All wavelengths and redshifts are in vacuum and are corrected to a heliocentric velocity standard. 
All distances are physical.
All stated errors are $1\sigma$, unless otherwise noted.

\section{MUSE and UVES observations}
\label{sec:data}
The data we use in this study are part of a MUSE GTO program aiming to study gas flows around galaxies using background quasars.
This  MusE GAs FLOw and Wind (MEGAFLOW) program 
 combines MUSE \citep{BaconR_10a}
and UVES \citep{DekkerH_00a} observations of 22 quasar fields  
and 
is described in \citet{SchroetterI_16a, ZablJ_19a, SchroetterI_19a}. The   quasar fields were selected based on the presence of at least $3$ strong ($W_r^{\lambda 2796}>0.5$--0.8\AA) \MgII\ absorption in SDSS spectra
and observed with MUSE with on average 3 hours of exposure time each. 
The observations and data reduction are described in \citet{SchroetterI_16a, ZablJ_19a}.

For each field, UVES follow up observations of the quasar have been carried out.
In this section, we present, first,  the detection of [\OII] emitters throughout all the 22 MUSE fields, and then the detection of \CIV\ and \MgII\ absorption systems in the UVES spectra.
We will combine those two samples in order to search for absorber counterparts to [\OII] emitters\footnote{Note, we use the same MUSE data reduction as used in \citet{SchroetterI_19a} and \citet{ZablJ_19a}.}.

\subsection{Galaxy detection in all MUSE fields from $z\geq1.0$}

The MUSE wavelength range spans from $\approx$4700 to $\approx$9300~\AA, {at a spectral resolution of R$\approx2000$ (or 150~\kms)}. 
This wavelength range allows {for} the detection of [\OII] emitters from redshifts 0.3 $\leq z \leq 1.5$.

{In order to find star-forming galaxies potentially associated with
\CIV{} or \MgII{} absorbers}, we  {searched} for [\OII] emitters in our MUSE fields, independently of any absorption systems,
but to ensure overlapping redshift ranges between \CIV\ and \MgII\ in UVES quasar spectra,
we restricted the search for  [\OII] emitters at redshifts 1.0--1.5, i.e in the 7400--9300~\AA\ MUSE wavelength range.
For this, we used narrow band images with the continuum subtracted throughout the cube wavelength range.
In the 22 MEGAFLOW fields, we found {215} [\OII] emitters and visually inspected each galaxy, making sure the [\OII] emission doublet was clearly identified {down to a flux detection limit of $10^{-17}$\flux}.
{In a forthcoming paper (Bouché et al. in prep.), we will characterize and present the detection completeness for the MEGAFLOW survey using a combination of automatic and visual inspection sources detection. 
Preliminary results based on inserting fake [\OII] emitters with realistic sizes and kinematic properties give a 50\%\ completeness level at $\approx8.0\times10^{-18}$ \flux\ (corresponding to a SFR of $\approx0.3$~\mpy at $z=1.0$) for a typical depth of 3 hr in the red part of the MUSE wavelength range {away from skylines}.}

\subsection{\CIV\ and \MgII\ absorption systems}

\begin{figure}
   \centering 
   \includegraphics[width=8.0cm]{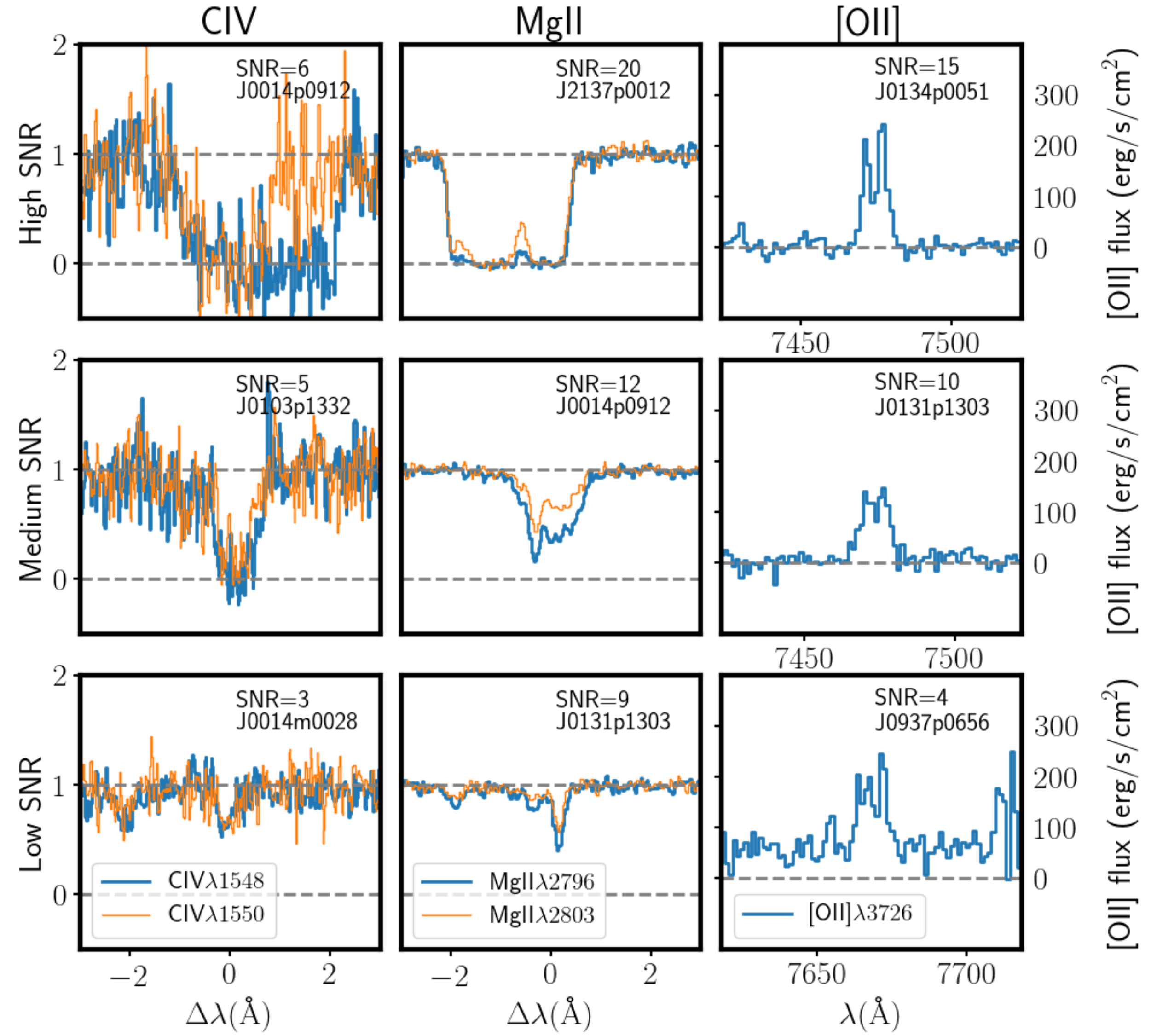} 
   \caption{Examples of \CIV\ (left) and \MgII\ (middle) absorption systems as well as [\OII] (right) emission for high, medium and low Signal to Noise Ratio (SNR) in the top, middle and bottom rows,  respectively.
   Each {panel} is independent and they were chosen arbitrary for examples purpose. 
   {Each SNR is shown for each panel as well as each field name.  The $x$-axis for [\OII] emission (right column) corresponds to observed wavelength.}
   } 
   \label{fig:snr_example} 
\end{figure}

Using the  MEGAFLOW UVES quasar spectra, we {searched} for \CIV\ and \MgII{} systems at $z=1$--1.55 , i.e. in the redshift range where both \CIV{} and \MgII{} are {covered in}  the UVES spectra and where [\OII] can be detected in the MUSE wavelength {range},
 down to a rest-frame equivalent width {(REW)} $W_r$ of $\sim$0.1\AA.
The detection limit for each field is based on the signal to noise ratio and we find an overall equivalent width limit EW$_{\rm lim}\approx$0.1~\AA\ for both \CIV\ and \MgII\ {in the observed frame}. 
Since all the absorption systems have $z>1$, the rest EW$_{\rm lim}$  is thus divided by 1+$z$, leading to REW$_{\rm lim}$=0.05~\AA. 
{Examples of absorption and emission systems are shown in Figure~\ref{fig:snr_example}. 
This Figure shows a selected sample of \CIV, \MgII\ and [\OII] systems detected at high, medium and low SNR. 
We emphasize here that each panel of this Figure is an independent system (different field,  redshift or position).}

\begin{figure}
   \centering 
   \includegraphics[width=8.0cm]{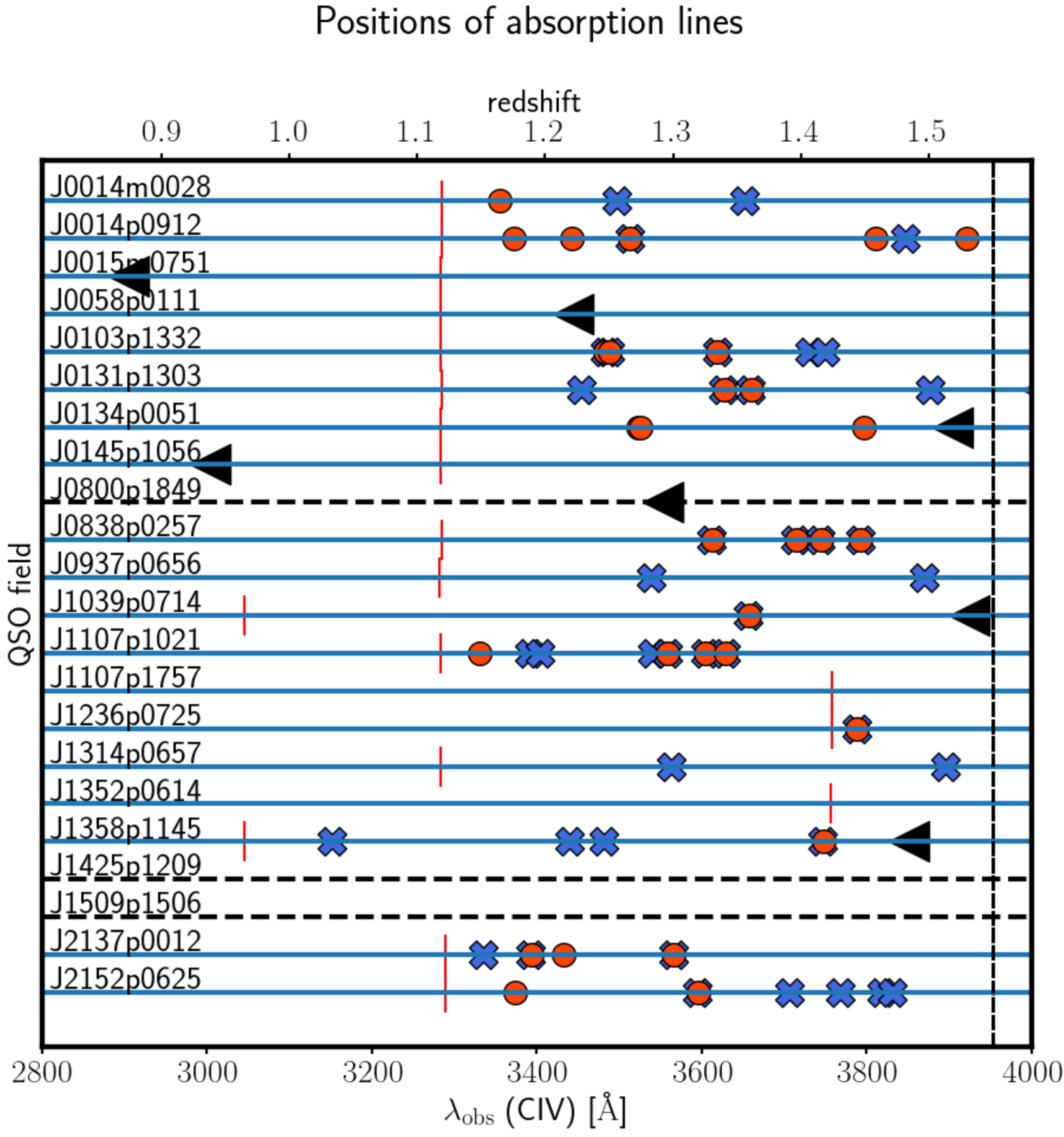} 
   \caption{ Observed wavelengths of \CIV\ systems in each MEGAFLOW QSO UVES spectra.
      The top $x$-axis represents the corresponding redshift.
   The blue crosses represent the  detected \CIV\ absorption system.
   The red circles represent the expected $\lambda_{\rm obs}(\CIV)$ for systems with detected \MgII{}. 
The red vertical lines represents the UVES lowest wavelength coverage and the black vertical line the $z=1.55$ redshift limit.
   The 3 horizontal black dashed lines are the fields for which the lowest UVES wavelength is above 4000~\AA\ {where \CIV\ is not covered at $z\approx$1-1.5}.
   The left black triangles show the quasar redshift positions with $z_{\rm QSO}<1.5$.
   } 
   \label{fig:sample} 
\end{figure}

We visually {searched} all the 22 QSO's UVES spectra for \CIV\ systems.
{UVES allows for \CIV\ absorption detection for redshift $z>1.0$ with a spectral resolution of R$\approx38000$\footnote{With a slit width of 1.2\arcsec\ and a CCD readout with 2x2 binning. 
UVES observation settings and data quality are detailed in \citet{ZablJ_19a} and \citet{SchroetterI_19a}. }}. 
{The UVES spectral coverage is continuous over the redshift range of $1.0 < z < 1.55$}.
Two of the authors performed this search independently and cross-checked the findings.
In short, what we call a "confident \CIV\ system" is a system having clear \CIV($\lambda \lambda$ 1548, 1550) absorption. 
Both doublet components {had to have} the same overall shape.
Figure~\ref{fig:sample} shows the redshifts of all the \CIV\ systems for each quasar sight-line.

In the 22 UVES spectra, we find a total of 41 \CIV\ absorption systems with redshifts between 1.0 and 1.55, for which we also have \MgII\ coverage.
Out of the 22 quasar fields, we only find \CIV\ absorption systems in 13 of them.
The 9 remaining fields either do not have UVES wavelength coverage (for 6 of them) or the quasars have lower redshift than 1.55 (for 3 of them).
It is worth mentioning that within our selected wavelength coverage from redshifts 1 to 1.55 (from $\approx 3100$~\AA~to $\approx 4000$~\AA), our UVES data do not have any spectral gap. 

\begin{figure}
   \centering 
   \includegraphics[width=8.0cm]{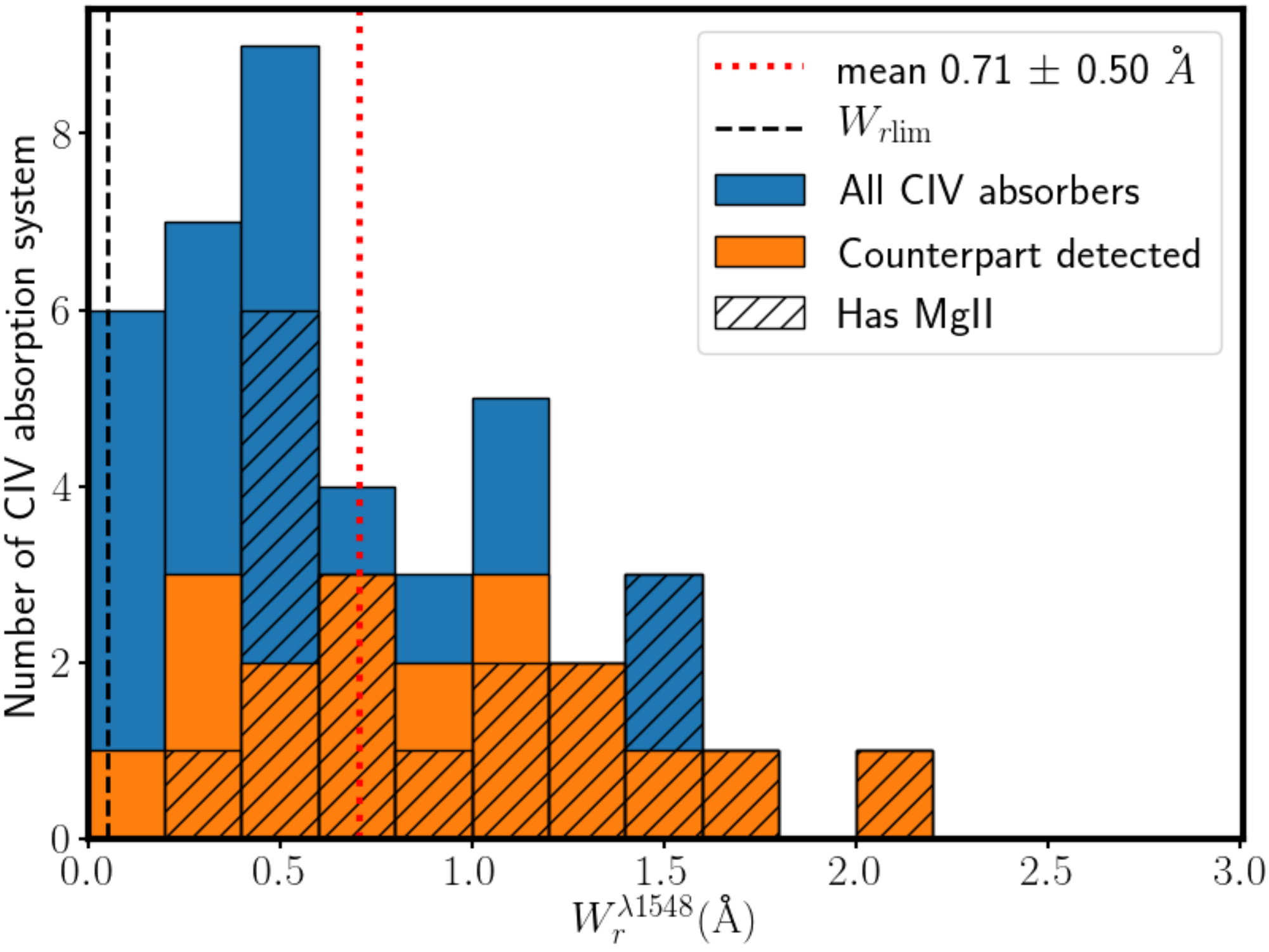} 
   \caption{Histogram showing number of \CIV\ absorption systems as a function of their REW for all the \CIV\ absorbers (in blue), the ones with galaxy counterparts detected (in orange) and the number of absorbers for which we also detect \MgII\, absorption lines (hatched). 
   The REW detection limit is shown by the black dashed vertical line} 
   \label{fig:detection_hist} 
\end{figure}

For those 41 \CIV\ absorption systems, we also {searched} for \MgII\ absorption lines.
We find that out of the 41 \CIV\ systems, {18} have \MgII\ absorption (see blue and hatched systems in Figure~\ref{fig:detection_hist}). 
We emphasize the fact that the MEGAFLOW survey was built based on \MgII\ absorption lines detected in quasar spectra from SDSS. 
Hence, we keep in mind that the numbers we give in this paper may be biased towards more \MgII\ absorption systems and that the MEGAFLOW selection may also increase the number of expected \CIV\ systems. 
However, since this survey has the advantage of having high-resolution quasar spectra for each of the MUSE fields, we begin with this biased sample before building larger ones. 

\begin{figure}
   \centering 
   \includegraphics[width=8.0cm]{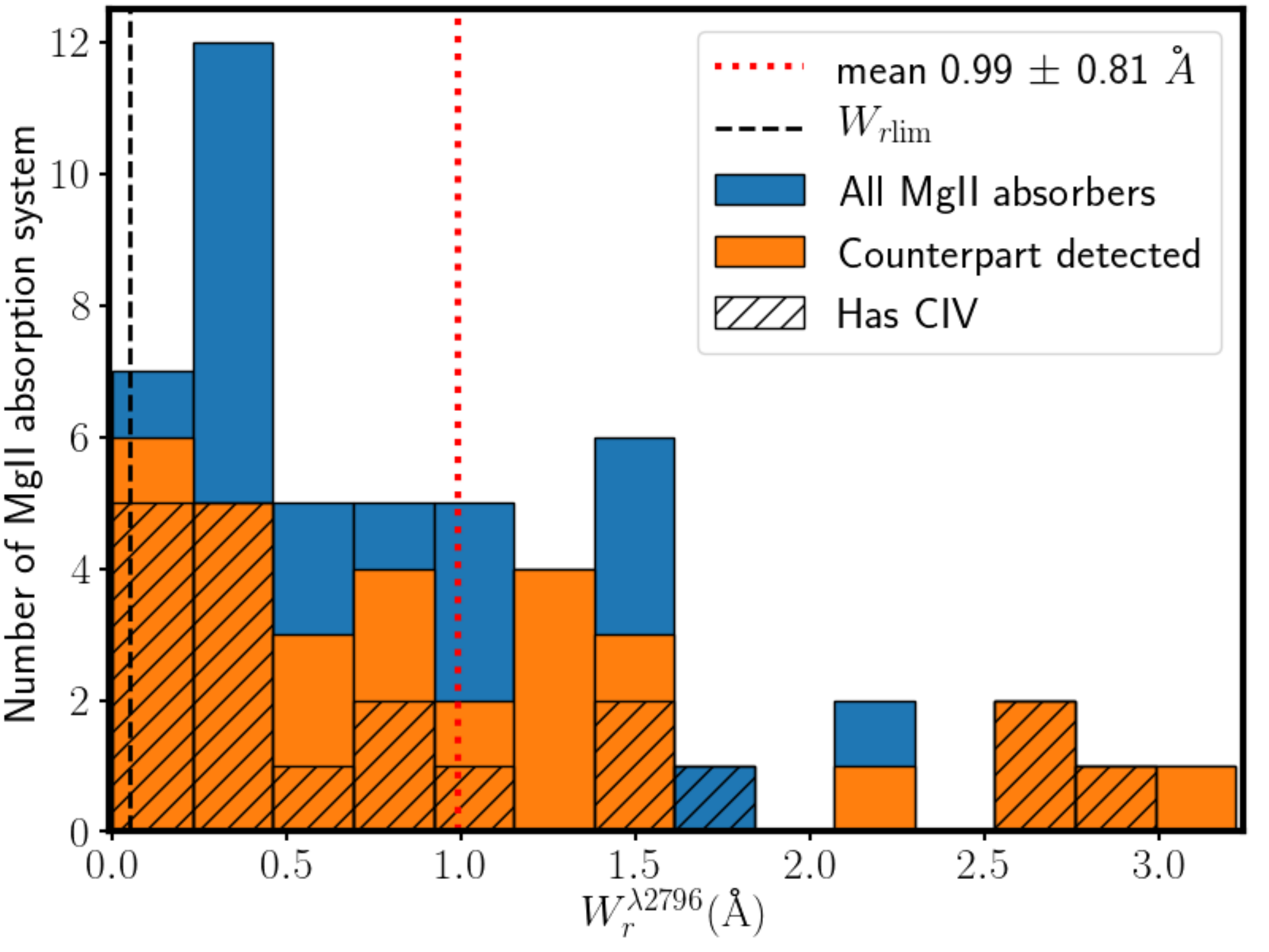} 
   \caption{$W_r^{\lambda 2796}$ distribution for all detected \MgII\ absorption systems.
}
   \label{fig:detection_hist_mgii} 
\end{figure}

In order to complete our analysis and for the purpose of the covering fraction study later in this paper, we also {searched} for all \MgII\ absorption systems in each of our UVES data, regardless of the presence of \CIV\ absorbers.
We first {searched} for any \MgII\ absorption system with redshift {between 1.0 and 1.55}. 
{We find 52 \MgII\ systems in total to begin with. 
Then, we rejected the ones for which we cannot observe \CIV\ absorption due to the UVES wavelength coverage, resulting in {30} \MgII\ systems where the \CIV\ feature is covered.
Out of these {30}, we {detected} \CIV\ in {18} systems.}
The distribution of REW for all the \MgII\ systems is represented in Figure~\ref{fig:detection_hist_mgii}.
Comparing the REW distribution of systems which have only \CIV\ and the ones with the presence of \MgII~(hashed histogram on Figure~\ref{fig:detection_hist_mgii}), we find that we are in good agreement with the study of \citet{SteidelC_92a} where they find that \CIV-only systems are peaked towards lower REW.
{This difference could come from the fact that \CIV\ may be more diffuse than \MgII\ around galaxies.}

Table~\ref{table:counterparts} summarizes all the different absorption systems for which  both absorption lines (\CIV\ and \MgII) can be detected.
Table~\ref{table:counterparts} lists the absorption system redshift with the corresponding REW of \CIV($\lambda \lambda$1548, 1550) and \MgII($\lambda \lambda 2796, 2803$) (if detected) absorption lines. 
{We note that each \CIV\ or \MgII\ doublet is resolved and each corresponding limit was applied to each individual doublet member.}
{Table~\ref{table:counterparts_mgii} shows \MgII\ absorption system properties for which \CIV{} is not covered in UVES.}

\begin{table}
\centering
\caption{Absorption system properties { (both \CIV{} and \MgII{} covered)}.}
\label{table:counterparts}
\begin{tabular}{clccccc}
\hline
Field name & $z_{\rm abs}$ & $W_r^{\lambda1548}$ & $W_r^{\lambda1550}$ & $W_r^{\lambda2796}$ & $W_r^{\lambda2802}$ & N$_c$\\
(1) & (2) & (3) & (4) & (5) & (6) & (7)\\
\hline
J0014m0028 & 1.16444 & $\leq0.05$ & $\leq0.05$ & 0.3251 & 0.2721 & 0\\
$\cdots$ & 1.25638 & 0.6108 & 0.3913 & $\leq0.05$ & $\leq0.05$ & 0 \\
$\cdots$ & 1.35566 & 0.0655 & 0.0762 & $\leq0.05$ & $\leq0.05$ & 0\\
\hline
J0014p0912 & 1.17550 & $\leq0.05$ & $\leq0.05$ & 0.9277 & 0.8453 & 0\\
$\cdots$ & 1.22115 & $\leq0.05$ & $\leq0.05$ & 1.2521 & 0.9717 & 5\\
$\cdots$ & 1.26575 & 1.5459 & 0.9266 & 0.3772 & 0.1909 & 0\\
$\cdots$ & 1.45844 & $\leq0.05$ & $\leq0.05$ & 0.2831 & 0.1541 & 0\\
$\cdots$ & 1.48212 & 0.1778 & 0.3613 & $\leq0.05$ & $\leq0.05$ & 0\\
\hline
J0103p1332 & 1.24646 & 0.5493 & 0.5053 & 0.2369 & 0.1577 & 0\\
$\cdots$ & 1.25029 & 0.5005 & 0.1065 & $\leq0.05$ & $\leq0.05$ & 0\\
$\cdots$ & 1.33451 & 0.4467 & 0.3516 & 0.3347 & 0.2231 & 0\\
$\cdots$ & 1.34053 & $\leq0.05$ & $\leq0.05$ & 0.1687 & 0.1107 & 1\\
$\cdots$ & 1.36112 & $\leq0.05$ & $\leq0.05$ & 0.3727 & 0.2989 & 2\\
$\cdots$ & 1.40648 & 0.3491 & 0.2141 & $\leq0.05$ & $\leq0.05$ & 0\\
$\cdots$ & 1.41847 & 0.9428 & 0.6973 & $\leq0.05$ & $\leq0.05$ & 0\\
\hline
J0131p1303 & 1.22845 & 0.2819 & 0.2384 & $\leq0.05$ & $\leq0.05$ & 0\\
$\cdots$ & 1.33945 & 0.7591 & 0.5660 & 0.1567 & 0.1024 & 1\\
$\cdots$ & 1.36075 & 0.3734 & 0.3213 & 0.3582 & 0.2862 & 2\\
$\cdots$ & 1.50130 & 0.0947 & 0.0695 & $\leq0.05$ & $\leq0.05$ & 0\\
\hline
J0134p0051 & 1.27246 & $\leq0.05$ & $\leq0.05$ & 0.3312 & 0.2847 & 0\\
$\cdots$ & 1.27411 & $\leq0.05$ & $\leq0.05$ & 0.3500 & 0.2266 & 0\\
$\cdots$ & 1.44917 & $\leq0.05$ & $\leq0.05$ & 0.4287 & 0.2362 & 1\\
\hline
J0838p0257 & 1.33042 & 1.4164 & 1.1392 & 1.5306 & 1.2020 & 1\\
$\cdots$ & 1.39601 & 0.5442 & 0.5272 & 0.0598 & 0.0228 & 1\\
$\cdots$ & 1.41490 & 0.7320 & 0.3133 & 0.8686 & 0.6654 & 2\\
$\cdots$ & 1.44629 & 1.4876 & 1.2233 & 0.7595 & 0.5980 & 0\\
\hline
J0937p0656 & 1.28260 & 1.0395 & 0.7001 & $\leq0.05$ & $\leq0.05$ & 0\\
$\cdots$ & 1.49647 & 1.0355 & 1.0112 & $\leq0.05$ & $\leq0.05$ & 0\\
\hline
J1039p0714 & 1.35888 & 1.1127 & 1.0100 & 2.4970 & 2.2989 & 3\\
\hline
J1107p1021 & 1.14907 & $\leq0.05$ & $\leq0.05$ & 0.7981 & 0.515 & 0\\
$\cdots$ & 1.18750 & 0.2123 & 0.1948 & $\leq0.05$ & $\leq0.05$ & 2\\
$\cdots$ & 1.19557 & 0.4752 & 0.1473 & $\leq0.05$ & $\leq0.05$ & 0\\
$\cdots$ & 1.28363 & 0.2810 & 0.1408 & $\leq0.05$ & $\leq0.05$ & 2\\
$\cdots$ & 1.29562 & 2.1644 & 1.9257 & 0.5323 & 0.2984 & 6\\
$\cdots$ & 1.32523 & 1.6618 & 1.4281 & 2.7651 & 2.6038 & 3\\
$\cdots$ & 1.34105 & 1.2897 & 1.0869 & 0.1009 & 0.0524 & 3\\
\hline
J1236p0725 & 1.44338 & 0.4736 & 0.3960 & 1.7089 & 1.5374 & 0\\
\hline
J1314p0657 & 1.29867 & 0.2399 & 0.1287 & $\leq0.05$ & $\leq0.05$ & 0\\
$\cdots$ & 1.51345 & 0.1353 & 0.1080 & $\leq0.05$ & $\leq0.05$ & 0\\
\hline
J1358p1145 & 1.03320 & 1.1543 & 0.9007 & $\leq0.05$ & $\leq0.05$ & 1\\
$\cdots$ & 1.21902 & 0.1625 & 0.1520 & $\leq0.05$ & $\leq0.05$ & 2\\
$\cdots$ & 1.24627 & 0.0615 & 0.0350 & $\leq0.05$ & $\leq0.05$ & 0\\
$\cdots$ & 1.41709 & 0.9830 & 0.7056 & 2.5651 & 2.3938 & 1\\
\hline
J2137p0012 & 1.15191 & 0.2559 & 0.1254 & $\leq0.05$ & $\leq0.05$ & 0\\
$\cdots$ & 1.18860 & 0.6404 & 0.4889 & 0.2686 & 0.1671 & 2\\
$\cdots$ & 1.21472 & $\leq0.05$ & $\leq0.05$ & 1.1219 & 1.0573 & 4\\
$\cdots$ & 1.30023 & 0.4736 & 0.1542 & 0.0731 & 0.0590 & 1\\
\hline
J2152p0625 & 1.17651 & $\leq0.05$ & $\leq0.05$ & 0.496 & 0.3924 & 0\\
$\cdots$ & 1.31870 & 1.3851 & 0.9527 & 1.3841 & 1.1038 & 2\\
$\cdots$ & 1.39149 & 0.8075 & 0.7151 & $\leq0.05$ & $\leq0.05$ & 1\\
$\cdots$ & 1.43101 & 1.0550 & 0.9595 & 0.9884$^\dagger$ & $\leq0.05$ & 3\\
$\cdots$ & 1.46349 & 0.4458 & 0.1634 & $\leq0.05$ & $\leq0.05$ & 0\\
$\cdots$ & 1.47132 & 0.5880 & 0.4002 & $\leq0.05$ & $\leq0.05$ & 0\\

\hline
\end{tabular} \\
1: Field name;
2: Absorption system redshift;
3: \CIV~($\lambda1548$) REW (\AA);
4: \CIV~($\lambda1550$) REW (\AA);
5: \MgII~($\lambda2796$) REW (\AA);
6: \MgII~($\lambda2802$) REW (\AA);
{7: Number of detected counterparts.}
Typical EW 1-$\sigma$ errors are 0.01~\AA. 
$\dagger$: this system is not considered as a detection since we do not detect the $\lambda 2802$ component which falls in the UVES coverage gap.
\end{table}

\begin{table}
\centering
\caption{\MgII\ absorption system properties (\CIV{} not covered).}
\label{table:counterparts_mgii}
\begin{tabular}{clccc}
\hline
Field name & $z_{\rm abs}$ & $W_r^{\lambda2796}$ & $W_r^{\lambda2802}$ & N$_c$\\
(1) & (2) & (3) & (4) & (5) \\
\hline
J0014m0028 & 1.05265 & 2.1232 & 1.7896 & 1 \\
\hline
J0058p0111 & 1.06528 & 1.5822 & 1.3268 & 0 \\
\hline
J0103p1332 & 1.04836 & 2.9956 & 2.7317 & 2 \\
\hline
J0134p0051 & 1.07010 & 0.1906 & 0.0534 & 2 \\
\hline
J0131p1303 & 1.01045 & 1.3197 & 0.9514 & 2 \\
$\cdots$   & 1.10436 & 1.0906 & 0.8421 & 2 \\
\hline
J0800p1849 & 1.12153 & 0.5252 & 0.5156 & 0 \\
\hline
J0838p0257 & 1.09958 & 0.1474 & 0.0538 & 1 \\
\hline
J1107p1021 & 1.01464 & 0.6477 & 0.5716 & 0 \\
$\cdots$   & 1.01643 & 1.1665 & 1.0339 & 4 \\
$\cdots$   & 1.04815 & 0.4274 & 0.3120 & 1 \\
\hline
J1107p1757 & 1.04801 & 2.1392 & 1.6854 & 0 \\
$\cdots$   & 1.06296 & 3.2741 & 2.6220 & 1 \\
$\cdots$   & 1.16320 & 1.5809 & 1.4023 & 1 \\
$\cdots$   & 1.32678 & 0.7279 & 0.5251 & 0 \\
$\cdots$   & 1.33079 & 0.9160 & 0.7724 & 2 \\
\hline
J1236p0725 & 1.33128 & 0.3908 & 0.2291 & 3 \\
\hline
J1352p0614 & 1.13736 & 1.4189 & 1.0794 & 0 \\
$\cdots$   & 1.20049 & 1.0326 & 0.9159 & 0 \\
\hline
J1509p1506 & 1.04659 & 1.5000 & 1.3575 & 0 \\
\hline
J2137p0012 & 1.04423 & 0.8518 & 0.8741 & 2 \\
\hline
J2152p0625 & 1.05366 & 0.5688 & 0.4714 & 2 \\
\hline
\end{tabular} \\
1: Field name;
2: Absorption system redshift;
3: \MgII~($\lambda2796$) REW (\AA);
4: \MgII~($\lambda2802$) REW (\AA);
{5: Number of detected counterparts.}
Typical EW 1-$\sigma$ errors are 0.01~\AA. 
\end{table}

Having independently detected the absorption systems in UVES spectra and the [\OII] emitters in MUSE, we next try to match galaxies to absorption systems.

\subsection{Absorber counterparts}

We now look for galaxies which possibly correspond to the \CIV\ absorption systems. 
For each detected galaxy, we {measured} the velocity difference\footnote{The velocity difference between the absorber redshift and the galaxy systemic redshift.} between \CIV\ absorption and [\OII] emission.
For the purpose of this work, a galaxy {was} considered to be a counterpart if $\vert \Delta v\vert <500$~\kms\footnote{We also {searched} for systems with $\Delta v<1000$~\kms\ but found no more associations.}.
With this criterion, we find 39 counterpart galaxies for 19 \CIV\ absorption systems, out to 250~kpc from the quasar.
Out of those 19 systems, {13} ($\approx{70\%}$) have \MgII\ absorption lines. 

One of the most interesting results is that out of all the {22} systems with only \CIV\ absorption detected (not \MgII), we detect counterparts for only {6} systems. 
This means that we do not detect any counterpart for $\approx${73}$\%$ of those "\CIV\ only" systems. 
This result suggests that the \CIV\ gas is probably located in the IGM rather than in the CGM or that \CIV\ is associated with  galaxies with lower SFR than \MgII, i.e with SFR$\lesssim0.3$~\mpy.
As already argued by \citet{RahmaniH_18}, a system having both \CIV\ and \MgII\ absorption lines appears to have a higher chance of a  counterpart detection, compared to a system with only \CIV\ absorption. 
Here we can confirm this statement as {70}$\%$ of our systems for which we find at least 1 counterpart have both \CIV\ and \MgII\ absorption lines. 

Using the same criterion to identify counterparts of {the 52 detected} \MgII\ systems, we find at least one galaxy counterpart for {32}~\footnote{Note that 21 of those {32} systems were part of the original MEGAFLOW selection. } systems ($\approx 63\%$~\footnote{Note that this is {the} detection rate for $z>1$ galaxies, compared to the $\approx{70\%}$ detection of the whole MEGAFLOW survey for redshifts $0.4<z<1.5$}), within 250~kpc from the quasar.
{On both Table~\ref{table:counterparts} and~\ref{table:counterparts_mgii}, the last column shows the number of galaxy counterpart detected for each system.}


\section{Measuring covering fractions}
\label{sec:covering}
\label{sec:methodo}

With our sample of [\OII{}] galaxies, we investigate the covering fractions of \CIV\ and \MgII\ at redshifts $1.0<z<1.5$.
{The covering fraction $\fc$ describes the probability of having an absorber detected (1) or not (0) at a distance $r$ from a galaxy, which is inherently  dichotomous. }
{$\fc$ can be estimated from the ratio between the number of galaxies at a given impact parameter $r\pm\Delta r$ which have an absorber detected (above some detection limit)  to the total number of galaxies $N_{\rm tot}$ at that impact parameter with or without a detected absorber. If `D' (`U') represent the detected (undetected) absorbers,  respectively, $f_{\rm c}(r)\equiv {D}/{D+U}$. 
In the notation of \citet{BordoloiR_14b}, $D$ is $N_{W_{\rm r}>W_{\rm r,lim}}(r)$, and $\fc$ is}
\begin{equation}
    f_{\rm c}(r)\equiv 
    \frac{N_{W_{\rm r}>W_{\rm r,lim}}(r)}
    {N_{\rm tot}(r)}
\end{equation}

In  general, the probability $p$ for { an absorber} to be detected (i.e., $Y=1$) can be described by any continuous function $p=L(t)$. Here, we use a logistic function~\footnote{See \citet{WildeM_21a} for another physical parameterisation.} for the probability $p$ to detect an { absorber}. The logistic function yields a continuous transition of p from 0 to 1:
\begin{equation}
p(Y=1)=\frac{1}{1+\exp(-t)}\equiv L(t)\label{eq:logistic}
\end{equation}
where $t$ is usually taken to be a linear combination of the independent variables, $X_n$, i.e., $t=\alpha+\beta_1 X_1+\cdots+\beta_n X_n$.\footnote{This  assumes that all variables are independent of one another, but covariance terms can be included.} Here, the {in}dependent variables will be impact parameter $r$ and redshift $z$.

{It is customary to determine
 $f_{\rm c}(r)$ in radial bins, by binning the data. However, this technique requires samples with large number of galaxies in each {bin} and thus can lead to large statistical uncertainties when the data {do} not populate each bin equally. 
 Here, we use a novel unbinned approach to
 fit the probability function $p$ directly using  a logistic function  in a Bayesian framework.}
As discussed in \citet{Hosmer2000}, this is commonly used to examine the possible relationship between a dichotomous variable (here whether the absorption are detected in the quasar spectra) and  independent variables ($X_n$, such as impact parameter, redshift, etc.). {This technique has several advantages, namely it uses all the information contained in the data even in {bins} where there would be few galaxies{.} 
The Bayesian framework is natural for this problem and provides also a more robust fitting procedure against outliers which could skew the mean of a particular bin.}

{A series outcomes $O$  can be generated by the Bernouilli distribution ${\rm Bern}(p)$,} since the observables are dichotomous, with values at 0 or 1, for undetected and detected systems, respectively. In summary, the  model is
\begin{eqnarray}
t&=&f(X_i;\theta)\label{eq:model}\\
p&=&L(t) \label{eq:model:p}\\
O&\sim&{\rm Bern}(p) \label{eq:model:logistic}
\end{eqnarray}
where $L(t)$ is given by Eq.~\ref{eq:logistic}, $f(X_i;\theta)$ linearly combines the data and map it to $L(t)$,  $X_i$ are the independent variables, $\theta$ are the model parameters, and $O$ are the simulated observables which can be compared to the observed data $Y_i$.

For the model function $t$, we choose a simple linear function $f$ of impact parameter $b$,
\begin{eqnarray}
f(b;\theta)&=&A \,(\log b-{C})\label{eq:model1}
\end{eqnarray}
where $A$ describes the slope of the covering fraction and $C$ the zero-point at 50\%\ covering fraction $f_{\rm c}=0.5$, since $L(0)=0.5$.  One could add a redshift dependent zero-point
\begin{eqnarray}
f(b,z;\theta)&=&A \,(\log b - {\rm ZP}(z))\label{eq:model2}
\end{eqnarray}
where  ${\rm ZP}(z) = B\,\log(1+z)+C$.
This second model is discussed/shown in \S~\ref{sec:results}.

We  used a Markov Chain Monte Carlo  
(MCMC) algorithm to estimate the best fitting parameters $\hat \theta$ for our model.  Because traditional MCMC algorithms are somewhat sensitive to the step size  and the desired number of steps, in what follows we use the No-U Turn Sampler (NUTS) of \citet{Hoffman2014} implemented in \textsc{Pymc3} \citep{pymc3},  a self-tuning variant of Hamiltonian Monte Carlo (HMC).
We typically used 2  MCMC chains per run and 4,500 iterations  per chain.

We stress that  the purpose of this exercise is to demonstrate the viability of our new technique for determining the covering fractions of metal lines in the regime of small number statistics.

\section{Results}
\label{sec:results}

\subsection{\MgII{} covering fraction}

The distribution of \MgII\  REWs {for all detected absorbers} is shown in Fig.~\ref{fig:detection_hist_mgii}. 
The \MgII\ systems have REWs of [0.1-2.5\AA] with a mean of $W^{2796}_r=\mgiimean\pm 0.8$\AA. 
{As discussed in Appendix~A2,
the mean REW $\langle W_r^{\lambda 2796}\rangle$ of our sample is biased towards high $W_r^{\lambda 2796}$ compared to random field \MgII{} absorbers  due to the MEGAFLOW survey selection criteria (\S~\ref{sec:data}).} The mean value for absorbers in the original MEGAFLOW selection is $\langle W_r^{\lambda 2796}\rangle=1.41$~\AA.

\begin{figure}
\centering
   \includegraphics[width=8cm]{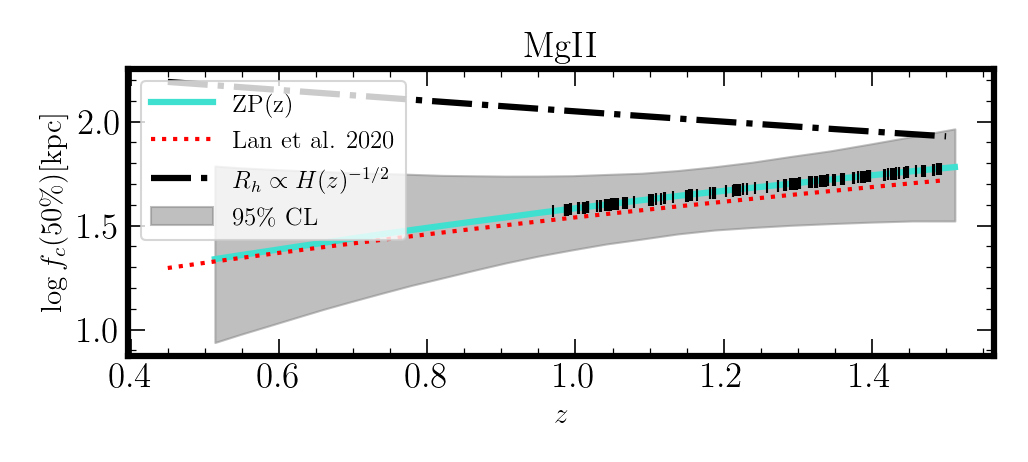} 
   \includegraphics[width=8cm]{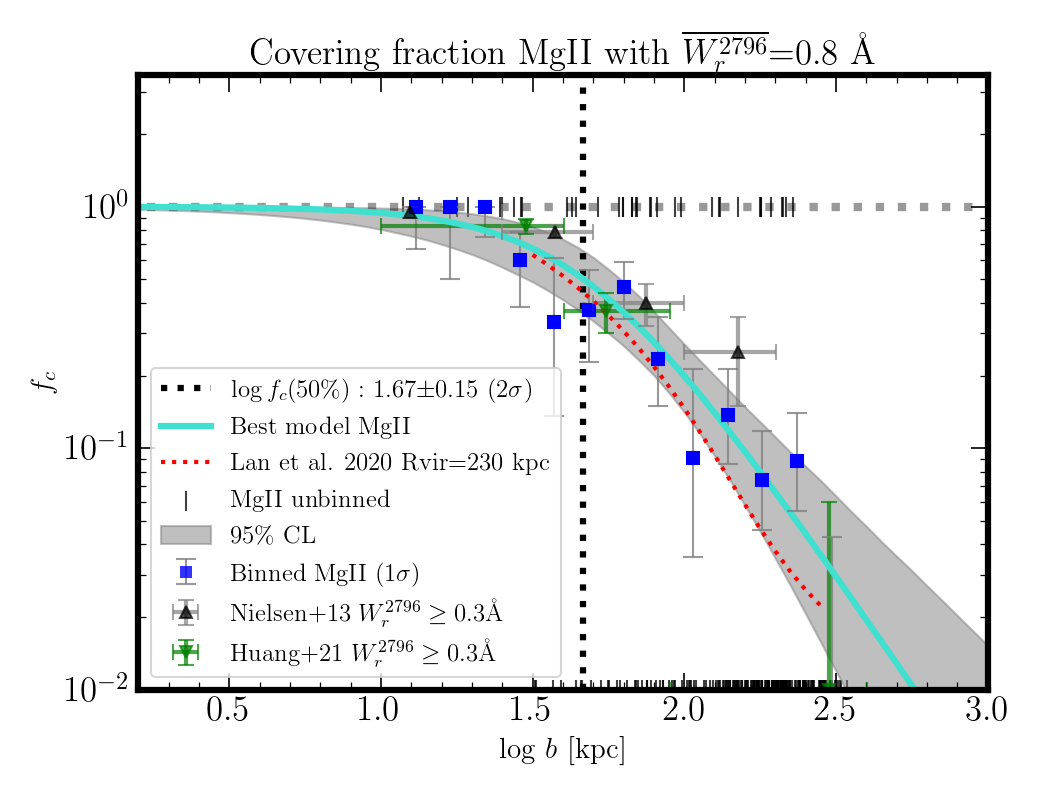}
\caption{{\it Bottom:} \MgII\ covering fraction  $f_{\rm c}$ as a function of impact parameter $b$.
The  fit to the {\it unbinned} data { (black ticks)} is shown (for $z=1.2$) with the solid blue line and its 2$\sigma$ confidence region is shown by the shaded region. The 50\%\ covering fraction is found to be at $\log b/\rm kpc\simeq1.7$ or $b\simeq50$~kpc.  
The dotted red line represents the fit found by \citet{LanT_19b} for $z=1.2$, $\log(M_\star/\rm M_\odot)=10$ and $R_{\rm vir}=250$ kpc.
{The solid black (green) triangles represent the \citet{NielsenN_13b} \citep{HuangY_21a} results, respectively, both at $z\sim0.5$.}
The solid blue squares with error bars show the binned data with $1\sigma$  { confidence intervals for binomial proportions (see text).}
The error bars are shown only with at least 2 galaxies contributing. 
  {\it Top:} The redshift evolution of the 50\%\ covering fraction with the $2\sigma$ predictive interval shown in grey. 
The redshift evolution found by \citet{LanT_19b} is shown with the dotted red line.
{ The dotted-dashed line shows the expected redshift evolution of halo sizes \citep[e.g.][]{MoH_98a}.}
}
\label{fig:fcovering:MgII}
\end{figure}

\Fig{fig:fcovering:MgII}(bottom) shows the \MgII\ covering fraction $f_{\rm c}$ as a function of impact parameter $b$. The solid line represents a logistic fit with our model (\Eq{eq:model2}) to the unbinned data (black ticks) with the 95\%\ confidence interval (grey area) derived from the Markov chain.
The 50\%\ covering fraction $f_{\rm c}(50\%)$ occurs at $\log b/\rm kpc=1.66\pm0.15$ ($2\sigma$), corresponding to 46$_{-13}^{+18}$~\kpc.
For comparison,  the filled squares with  error bars  represent the binned data. These error bars
are $1\sigma$ confidence intervals { computed for proportion of a binomial distribution  \citep{WilsonE_27a,CameronE_11a}.}
Comparing these $1\sigma$ errors on the binned data to the 95\%\ confidence interval to the unbinned data, this figure shows  the benefit of a parametric fit to the unbinned data. {The fitted parameters for the \MgII{} covering fractions are listed in Table~\ref{tab:my_label} with (Eq.~\ref{eq:model2})
and without the redshift evolution term (Eq.~\ref{eq:model1}).}

\Fig{fig:fcovering:MgII}(top) shows the redshift evolution determined from our model fit (\Eq{eq:model2}). The solid line shows  that $f_{\rm c}(50\%)$ evolves as $(1+z)^{B}$ with $B=1.96^{+2.32}_{-2.68}$ (2$\sigma$).
The dotted-dashed line represents the redshift evolution of dark-matter halos $R_{\rm vir}\propto H(z)^{-1}$ \citep[e.g.][]{MoH_98a} of a given mass.
Comparing the dotted-dashed line to the redshift evolution of $f_{\rm c}$, one sees that the \MgII{} halo is becoming proportionally smaller.

Our results are broadly in agreement with the study of \citet{NielsenN_13b} (grey triangles) which used a sample of 182 galaxies towards 134 quasar sight-lines over a wide redshift range of 0.07$\leq z \leq 1.12$. Among other literature results of \MgII{} covering fractions,
 \citet{LanT_19b}  stands out with a sample of 15,000 \MgII{}-galaxy pairs in SDSS with $W^{2796}_r>0.4~\AA$ (albeit with photometric redshifts).
 Their  $f_{\rm c}$ as a function of radius (redshift) is shown with the dotted line in the bottom (top) panel of Fig.~\ref{fig:fcovering:MgII}, respectively. In spite of our much smaller sample, our results are in good agreement with theirs. 
 The dotted line  in \Fig{fig:fcovering:MgII}(bottom) from \citet{LanT_19b} has been calculated for absorbers with $W_r^{2796}\geq1$\AA~\footnote{\citet{LanT_19b} does not provide a fit for $W_r^{2796}>0.4$\AA{} absorbers, but the equivalent width dependence seems small compared to the other parameters.}, at $z=1.2$, for star-forming galaxies with $\log(M_\star/M_\odot)=10.0$ assuming a $\Rvir\approx200-250$~\kpc{} from their Eqs 7--9.  The { dotted} line in \Fig{fig:fcovering:MgII}(top) has been calculated by solving numerically for the redshift evolution of $f_{\rm c}$ in \citet{LanT_19b}. { 
 Overall, our  results, from a much smaller sample, are very similar to those of \citet{LanT_19b}   who used 15,000 pairs.}
 
 It is interesting to compare the halo mass for \MgII\ absorbers  corresponding to $\Rvir\approx200-250$~\kpc, 
 namely $\log \Mh/\rm M_\odot\approx12.6$, to
 the mass scale  obtained by \citet{BoucheN_06c} from a  clustering analysis of 2,500 $z=0.5$ absorbers \citep[see also][]{GauthierJR_09a,LundgrenB_09a,LundgrenB_11a}:
 $\log \Mh/\rm M_\odot\approx 12.5\pm0.4$ for their absorbers with $0.3<{W_r^{2796}}<1.15$.  The covering fraction could also be
weakly dependent  on stellar mass \citep[e.g.][]{ChenHW_10a, NielsenN_13b,LanT_19b}, but we reserve such an analysis to a forthcoming paper on the full MEGAFLOW sample. Nonetheless, these results are a demonstration of the power of our fitting technique on unbinned data.

\subsection{\CIV\ covering fraction}

{In this section, we use the galaxies for which \CIV\ could be detected from our UVES coverage.
This lowers the number of [\OII] emitters from 215 to 141.}
From a total of {141}  [\OII] emitters in the MUSE data with redshift $>$1.0, 
after looking at the redshift differences between each emitter and \CIV\ absorption systems, we found {39 galaxy counterparts corresponding to 19 \CIV{} absorption systems.}
Incidentally, {70\%}\ ({13} out of 19) of those systems have detectable \MgII\ absorption lines as well~\footnote{MEGAFLOW was not designed around \CIV\ absorption system, this {70\%}\ might be indirectly biased by our \MgII\ selection.}.

\begin{figure}
\centering
   \includegraphics[width=8cm]{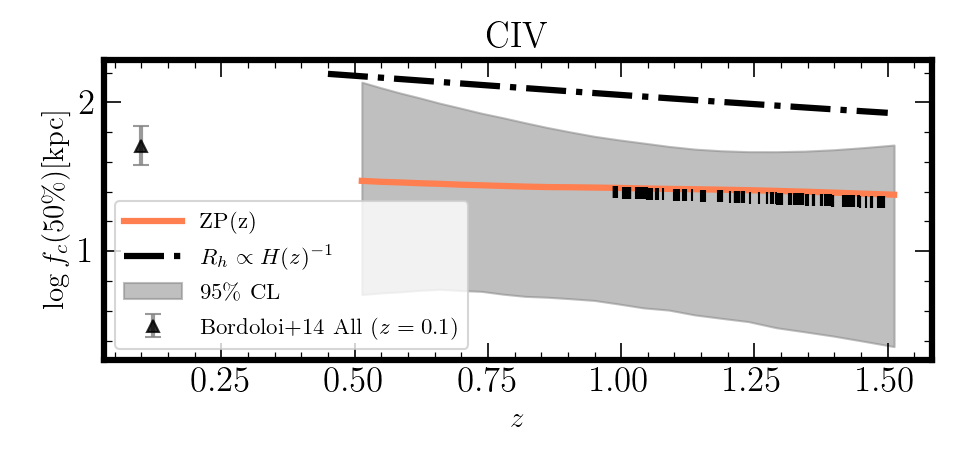} 
   \includegraphics[width=8cm]{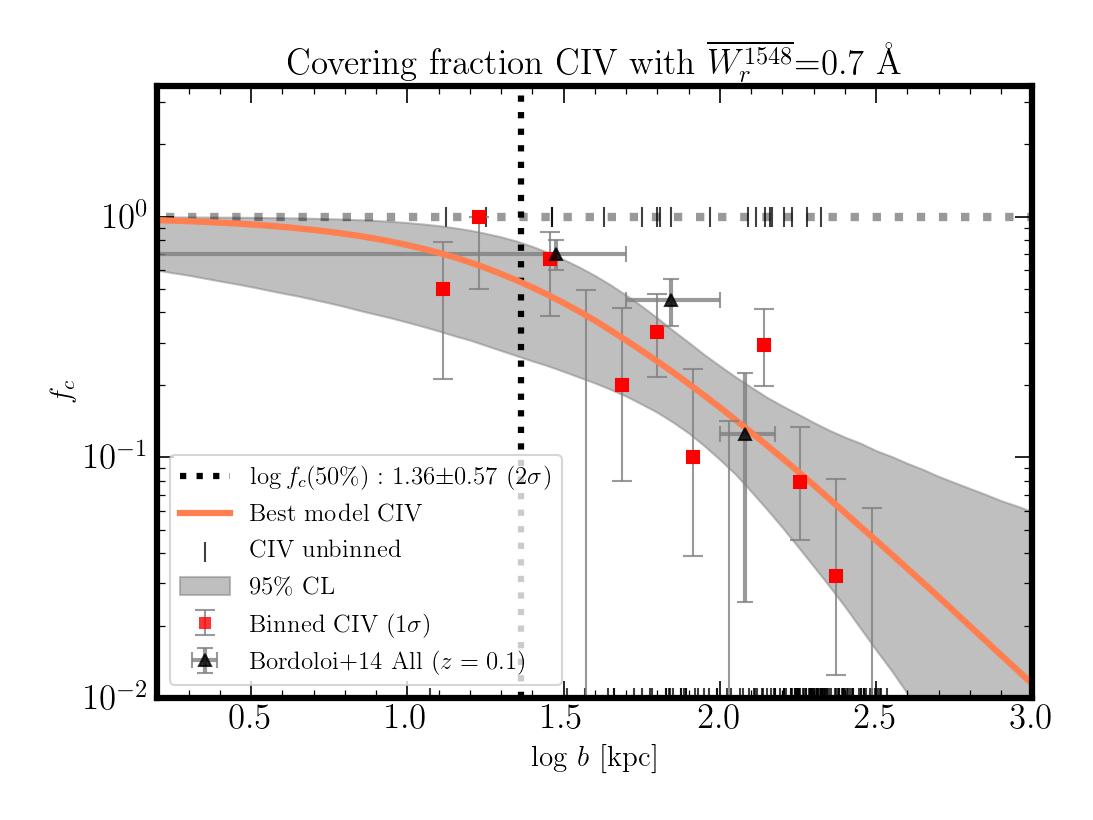}
\caption{{\it Bottom:} \CIV\ covering fraction  $f_{\rm c}$ as a function of impact parameter $b$.
 The  fit to the {\it unbinned} data { (black ticks)}  is shown (for $z=1.2$) with the solid orange line and its 95\%\ (2$\sigma$) confidence region is shown by the shaded region. {The black triangles represent the \citet{BordoloiR_14b} $z=0$ results.}
 The solid red squares with error bars show the binned  data with { $1\sigma$ confidence intervals for binomial proportions (see text)}.  
The 50\%\ covering fraction is found to be at $\log b/\rm kpc\simeq1.4$ or $b\simeq25$~kpc. 
{\it Top}: The redshift evolution of the 50\%\ covering fraction with the $2\sigma$ predictive interval shown in grey. { The dotted-dashed line shows the expected redshift evolution of halo sizes \citep[e.g.][]{MoH_98a}.}
}
\label{fig:fcovering:CIV}
\end{figure}

\Fig{fig:fcovering:CIV}(bottom) shows the covering fraction as a function of impact parameter $b$ { 
from our logistic fit (Eq.~\ref{eq:model2}) to the unbinned data (black ticks)}. The 50\%\ covering fraction occurs at $\log b/\rm kpc=$\fCIV ($2\sigma$), i.e. 23$^{+62}_{-16}$ \kpc.
For comparison, the filled squares with error bars 
represent the binned data. { These error bars
are  $1\sigma$ confidence intervals as in \S~4.1.}
Our $f_{\rm c}$(\CIV) appears to be in good agreement with the \citet{BordoloiR_14b} study (solid triangles), which used a sample of 43 low-mass $z\leq0.1$ galaxies around sight-lines with \CIV\ $W_{\rm r}^{1548}$ greater than $\sim0.1$\AA.

 {
The top panel of \Fig{fig:fcovering:CIV} shows the fitted redshift evolution with the 95\%\ predictive interval.}
{ The solid line represent the fitted $(1+z)^{B}$ evolution with $B=-0.7^{+3.9}_{-4.0}$ (2$\sigma$), i.e. consistent with no evolution. The tick marks show the location of the unbinned data.The fitted parameters for the \CIV{} covering fractions are listed in Table~\ref{tab:my_label} with (Eq.~\ref{eq:model2})
and without the redshift evolution term (Eq.~\ref{eq:model1}).
}
{The solid triangle in this figure represent the  \CIV\ covering fraction $f_{\rm c}(50\%)$ from the study of \citet{BordoloiR_14b} at $z=0.1$. This data point is obtained after applying our methodology to their 44 data points which yielded a size of $\log b$/kpc 1.71$\pm0.26$ (2$\sigma$) for the 50\%\ covering, i.e. 51$^{+42}_{-23}$ kpc.}

\begin{table*}
    \centering
    \begin{tabular}{cccccc}
    & \multicolumn{3}{c}{Fitted parameters}\\
          & A [95\%] & B [95\%] & C [95\%]  & $\log f_c$(50\%) & Model \\
         \hline	
\MgII & $-4.2$ [-5.7;-2.9] 
& 2.0 [-0.7; 4.3]
& 1.0 [0.2; 1.9] 
& \fMgII
& Eq.~\ref{eq:model2}
\\ 
\MgII & $-4.7$ [-6.3;-3.2]
& --
& 1.6 [1.5;1.8]
& 1.66$\pm0.15$
& Eq.~\ref{eq:model1}
\\
        \CIV 
& $-2.8$ [-4.5;-1.2] 
& -0.7 [-4.7;3.2]	
& 1.6 [0.3;2.8]  
& \fCIV
& Eq.~\ref{eq:model2}
\\
\CIV & $-3.0$ [-4.6;-1.6]
& --
& 1.4 [0.9;1.7]
& 1.44$\pm$0.39 
& Eq.~\ref{eq:model1}\\
        \hline
    \end{tabular}
    \caption{Fitted parameters ($A,B,C$) for $f_{\rm c}$ for \MgII{} and \CIV{} in relation to Eqs.~\ref{eq:model1}-\ref{eq:model2} along with their 95\%\ confidence interval. The $\log f_c$ column shows the distance at which $\fc$ reaches 50\%\ at $z=1.2$. 
    }
    \label{tab:my_label}
\end{table*}

\section{Discussion and Conclusions}
\label{sec:conclusion}

This study is divided into two parts: (i) the independent detection of absorption systems in MEGAFLOW UVES and [\OII] emitters in MEGAFLOW MUSE with the identification of galaxy counterparts to the absorbers; (ii) measurement of the covering fractions of \CIV\ and \MgII.

{We adopted an innovative Bayesian logistic regression and have shown its ability  }
(\S~\ref{sec:methodo})  to determine the covering fractions of absorption lines (such as \MgII, \CIV) with limited samples.  This approach allows one to fit a parametric model directly to unbinned binary data (1/0)  where 1 represents an absorption detection and 0 its absence.
The advantage of our method is that it does not suffer from the limitations inherent when using binned data.

Using the  MEGAFLOW UVES quasar spectra, we searched for \CIV, \MgII{} systems at $z=1$--1.5, i.e. in the redshift range where both \CIV{} and \MgII{} are covered by UVES and where [\OII] could be detected in the MUSE wavelength coverage,  down to a rest-frame equivalent width of 0.05\AA.
We {detected} a total of  {52} \MgII\ absorption systems and 41 \CIV\ absorption systems independently{, having large mean REWs of 0.99 and 0.71~\AA{} respectively}.
50\%\ and 60\%\ of the 41 \CIV\ and 32 \MgII\ systems\footnote{The 32 \MgII\ systems are the ones {with \CIV\ coverage.}}, respectively, have absorption by both ions.

Using the 22 MEGAFLOW MUSE fields, we searched for [\OII] emitters  at redshift $z=$1.0--1.5 and found a total of {215} emitters down to a flux limit of $\approx10^{-17}$ \flux\ \citep[][]{SchroetterI_16a, SchroetterI_19a, ZablJ_19a}. 
For each of these {215} [\OII] galaxies, we searched in our catalog of \CIV{} systems with REW $>0.05$~\AA\ for matches within 500 \kms\ and $b<250$ kpc. 
We found 39 galaxies associated with \Nciv\ \CIV{} systems.
For each galaxy, we also searched for corresponding \MgII{} systems and found 67 galaxies associated with {32} \MgII{} systems.

 We find that {70\%} of the systems for which we detect at least one galaxy counterpart have both \CIV\ and \MgII\ absorption lines. 
{Globally, these results point towards a physical picture where \MgII\ and \CIV\ ({with mean} REWs of $\approx$1\AA) are associated {with} individual L$_\star$/sub-L$_\star$ galaxies.}
We also find that {{73}\%} of the 'only \CIV' systems do not appear to have any galaxy counterpart in our MUSE fields.
This suggests that \CIV-only gas is more likely to be part of the IGM, i.e. beyond 250 kpc rather than of the CGM, or that \CIV\ is preferentially associated with very low mass galaxies (with SFR$\lesssim0.3$ \mpy).

Then, using the closest associated galaxies in our small sample of \Nciv\ ({32}) \CIV{} (\MgII{}) absorber-galaxy pairs, {we {studied} the
covering fraction and its redshift evolution using a logistic regression to the unbinned data in a Bayesian framework (\S~\ref{sec:methodo}). 
For \MgII{}, we find that $f_{\rm c}(r)$ reaches 50\%\ at  $r_{50}\approx50$~kpc ($46^{+{18}}_{-13}$; 2$\sigma$) (\Fig{fig:fcovering:MgII} bottom panel). Whereas for \CIV, we find that the size of the warm \CIV\ halo is somewhat smaller with $r_{50}\approx25$~kpc (23$_{-16}^{+62}$; $2\sigma$) (\Fig{fig:fcovering:CIV} bottom panel).  However, given the current sample size, this difference is not significant at $2\sigma$.}

Regarding the redshift evolution of $\fc(r,z)$ for strong \MgII{} absorbers with mean $W_r\simeq1$\AA\ (Fig.~\ref{fig:fcovering:MgII} top panel), { we found that $f_{\rm c}(50\%)$ evolves as $(1+z)^{B}$ with $B=1.96^{+2.32}_{-2.68}$.
Although the uncertainty of $B$ is large due to our small sample size, such that the relation is consistent with no redshift evolution, the best-fit $B$ value agrees with that derived by \citet{LanT_19b} $\propto(1+z)^{2}$ with a sample of 15,000 pairs.
Therefore, the two results combined support a physical picture where \MgII\ gas/halo decreases over time.} 
{We note that }whether this redshift index $B$ is $\sim0$ or 2, 
it appears that the \MgII\ gas/halo is becoming smaller with time {\it relative} to the dark-matter halo.

 Regarding the redshift evolution of $f_{\rm c}(r,z)$ for \CIV{} absorbers (Fig.~\ref{fig:fcovering:CIV} top panel), {our data over $z\sim1-1.5$ do not show significant redshift evolution.
 However, extending the redshift baseline to $z\sim0.1$ by including the \citet{BordoloiR_14b} data reveals that the \CIV{} covering fraction may in fact increase over time and possibly co-evolve with the dark-matter halo evolution  (dotted-dashed line in top panel).}
 This would indicate that the warm CGM probed by \CIV{} is growing with the halo size.
 This co-evolution is expected under a scenario where the warm gas is in hydrostatic equilibrium with the halo.

{With the final MEGAFLOW survey and upcoming larger surveys, it will be possible to address this redshift evolution and the potential dependence of covering fraction on azimuthal angle and/or EWs, thus yielding a better understanding of the CGM.}

\section*{Acknowledgments}
{We thank the referee for useful comments which led to an improved paper}.
This work received funding from the grants 3DGasFlows (ANR-17-CE31-0017), and the OCEVU Labex (ANR-11-LABX-0060) from the French National Research Agency (ANR).

{\it Software:} This work made use of the following open-source software: \textsc{Numpy} \citep{numpy}, \textsc{Scipy} \citep{scipy}, \textsc{Matplotlib} \citep{matplotlib}, and \textsc{Pymc3} \citep{pymc3}.
SM is supported by the Alexander von Humboldt-Stiftung via the Experienced Researchers fellowship. 

\section*{Data availability}
The data underlying this article are available in the ESO archive (http://archive.eso.org). The reduced data will be shared on reasonable request to the corresponding author.

\bibliographystyle{aa}
\bibliography{references}	
\label{lastpage}

\appendix
\section{}
\subsection{Additional Material}

\begin{figure}
   \centering 
    \includegraphics[width=8.0cm]{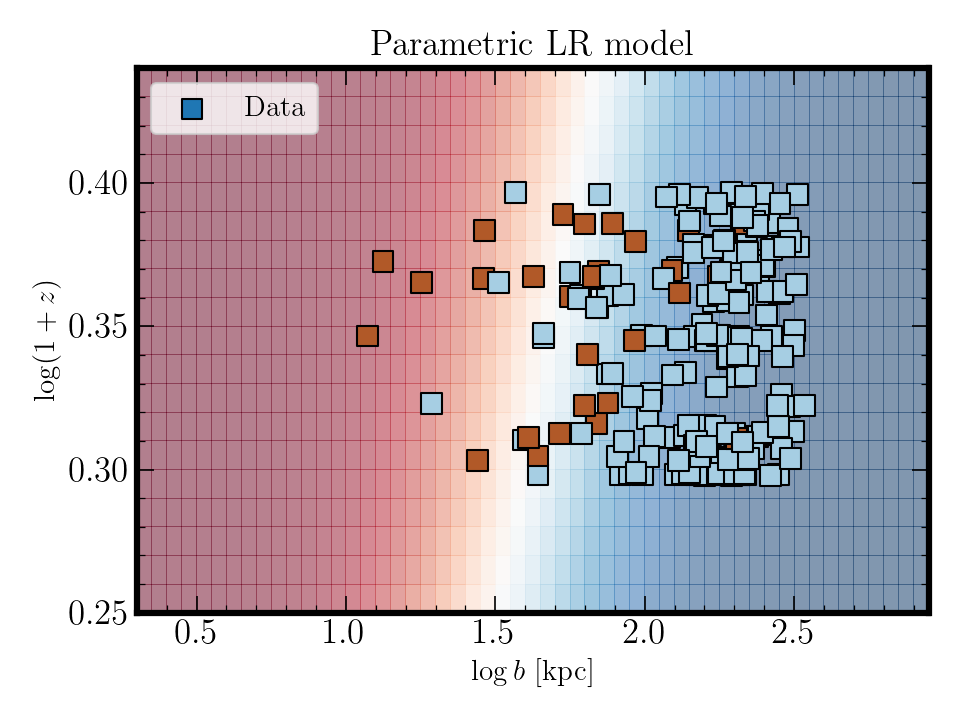} 
  \caption{ Two dimensional covering fraction of \MgII{} as a function of impact parameter $\log b$ and redshift $\log 1+z$. The unbinned data is shown as blue (red) squares  representing the undetected (detected) systems, respectively.  The \MgII{} covering fraction from the logistic fit is represented by the background color with low (high) covering fractions represented as blue (red) respectively.  } 
   \label{fig:fcovering:MgII:2D} 
\end{figure}

\begin{figure}
   \centering 
    \includegraphics[width=8.0cm]{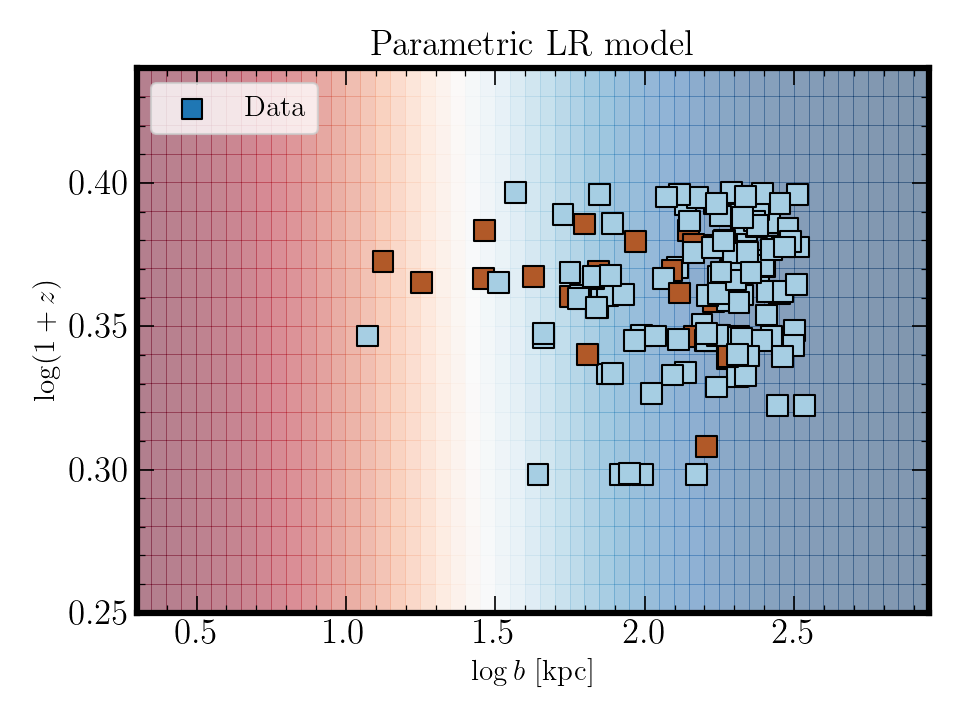} 
   \caption{Two dimensional covering fraction of \CIV{} as in \Fig{fig:fcovering:MgII:2D}.
   } 
   \label{fig:fcovering:CIV:2D} 
\end{figure}

{
\Fig{fig:fcovering:MgII:2D} shows two dimensional covering fraction of \MgII{} as a function of impact parameter $\log b$ and redshift $\log 1+z$ using the logistic regression described in \S~\ref{sec:methodo} (Eq.~\ref{eq:model2}). The fitted covering fraction is represented by the background color from low to high covering fractions represented as blue and red respectively.
}{
Similarly, \Fig{fig:fcovering:CIV:2D}  shows two dimensional covering fraction of \CIV{}.
 }
 
\subsection{Strong absorbers distribution}

\begin{figure}
   \centering 
       \includegraphics[width=8.0cm]{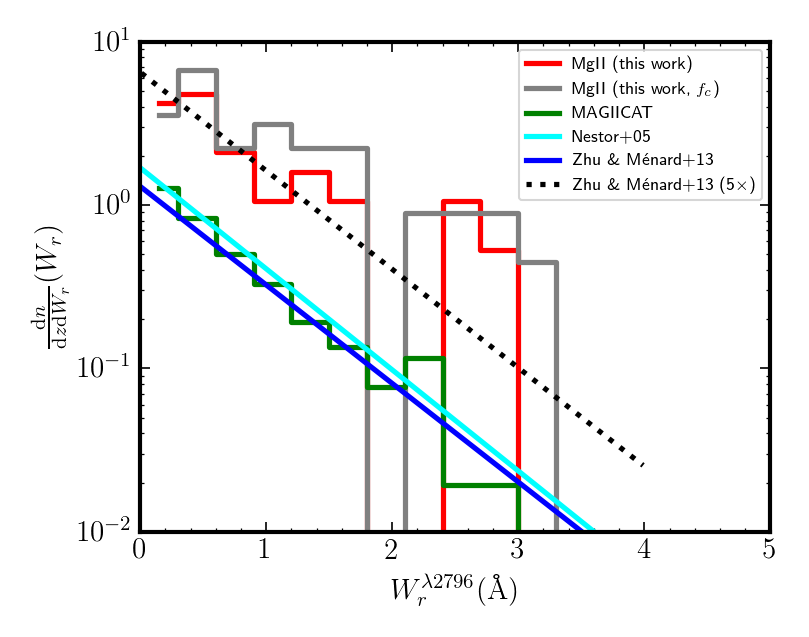}
   \caption{{Rest-frame $W_r^{\lambda 2796}$ distribution  of the 32 \MgII{} systems with \CIV\ (red histogram) and the 52 \MgII{} systems (grey histogram) using in \Fig{fig:fcovering:MgII}.
   The cyan (blue) curve represent the $W_r$ distribution from \citet{NestorD_05a} \citep{ZhuG_13a}
   at $z=1.2$, respectively.
   For comparison, the MAGIICAT data is shown with the green histogram \citep{NielsenN_13a}.
   }} 
   \label{fig:mgii_distrib} 
\end{figure}

In Figure~\ref{fig:mgii_distrib},
we compare the distribution of $W_r^{\lambda 2796}$ with literature samples. Our sample ({in red and grey}) has an overabundance of strong absorbers compared to other \MgII\ surveys due to the  MEGAFLOW selection of multiple strong \MgII\ absorbers in quasar spectra. 
However, the lower REW population ($\log(W_r^{\lambda 2796})\leq -0.5$) of MEGAFLOW absorbers found in this study, which are not pre-selected, are in agreement with other studies.

\end{document}